# Optimization of high performance spintronic terahertz sources

Fengwei Guo,[1] Xiaojun Wu,[1,*] Tianxiao Nie,[2,*] Chun Wang,[3,5] Shengyu Shan,[1] Chandan Pandey,[2] Meng Xiao,[1] Bo Wang,[3,5] Lianggong Wen,[4] Cunjun Ruan,[1] Jungang Miao,[1] Li Wang,[3] Yutong Li,[3,5] and Weisheng Zhao,[2,4]

[1]*Beijing Key Laboratory for Microwave Sensing and Security Applications, School of Electronic and Information Engineering, Beihang University, Beijing, 100083, China*
[2]*Fert Beijing Institute, BDBC, and School of Microelectronics, Beihang University, Beijing 100191, China*
[3]*Beijing National Laboratory for Condensed Matter Physics, Institute of Physics, Chinese Academy of Sciences, Beijing 100190, China*
[4]*Beihang-Goertek Joint Microelectronics Institute, Qingdao Research Institute, Beihang University, Qingdao, 266000, China*
[5]*School of Physical Sciences, University of Chinese Academy of Sciences, Beijing 100049, China*
*\*xiaojunwu@buaa.edu.cn, nietianxiao@buaa.edu.cn;*

To achieve high efficiency and good performance of spintronic terahertz sources, we propose and corroborate a remnant magnetization method to radiate continuously and stably terahertz pulses from W/CoFeB/Pt magnetic nanofilms without carrying magnets on the transmitters driven by nJ femtosecond laser pulses. We systematically investigate the influences of the pumping central wavelength and find out the optimal wavelength for a fixed sample thickness. We also optimize the incidence angle of the pumping laser and find the emission efficiency is enhanced under oblique incidence. Combing the aforementioned optimizations, we finally obtain comparable radiation efficiency and broadband spectra in W/CoFeB/Pt heterostructures compared with that from 1 mm thick ZnTe nonlinear crystals via optical rectification under the same pumping conditions of 100 fs pulse duration from a Ti:sapphire laser oscillator, which was not previously demonstrated under such pulse duration. We believe our observations not only benefit for a deep insight into the physics of femtosecond spin dynamics, but also help develop novel and cost-effective broadband spintronic terahertz emitters.

## I. Introduction

Femtosecond laser driven spintronic terahertz sources based on ferromagnetic/heavy metal (FM/HM) heterostructures have become one of the most promising candidates for the next-generation of novel terahertz emitters, which have attracted much attention these years [1–15]. Benefiting from strong absorption of subband electrons with asymmetry density distribution and different mobility of spin-up and spin-down [16], highly efficient heat-assisted spin converted charge photocurrents due to inverse spin Hall effect can radiate ultrafast terahertz pulses [1]. This kind terahertz source has been demonstrated to be very efficient, and can be comparable with nonlinear crystal or photoconductive antenna-based terahertz sources [2]. Furthermore, these magnetic terahertz sources have superiorities such as low cost, compactness and without direct current bias voltages, which make it very promising for commercial terahertz spectroscopy and imaging applications. Another unique property is that it can emit ultrabroadband terahertz radiation covering from 1-30 THz frequency range when driven by 10 fs laser oscillators due to the lack of phonon absorption limitation [2]. Compared with two-color pumped air plasma-based or transition radiation mechanism-based ultrabroadband terahertz sources [17,18], these magnetic nanofilm-based terahertz sources are all solid-state with very high stability and ease of use, while plasma-based sources require expensive femtosecond laser amplifier systems with sacrificing signal-to-noise ratio for ultrabroadband material recognitions. Beyond the numerous promising applications in the conventional weak-field terahertz systems, spintronic terahertz sources also have the feasibility to generate strong-field terahertz transients for nonlinear investigations [7]. Driven by a Ti:sapphire laser amplifier with 4 mJ power and, 40 fs duration, more than 300 kV/cm intense terahertz peak electric field has been achieved in W/CoFeB/Pt heterostructures [7]. Since this kind of emitter can be fabricated in wafer scale, it can also turn out to be one of the best candidates to fill the "new gap" of strong-field terahertz frequency region [19].

Since this kind of spintronic terahertz source has just emerged in 2013 [1], it is still in its infancy. The previous researches have been mainly concentrated on revealing the radiation mechanism or spin dynamics investigation employing the terahertz emission spectroscopy as a contactless, time-resolved, ultrafast Ampere meter [1]. However, recently, some works turn to study and develop this kind of promising source [2–14]. For conventional bilayer FM/HM heterostructures like Co/Pt, CoFeB/Pt and Fe/Pt, the back-flow spin current has been neglected without considering effective recycle and reuse [1]. However, when adding another capping layer with an opposite spin Hall angle like W [2], the ignored spin currents can be further converted to in-plane transverse charge currents, as a consequence, constructively boost the forward terahertz radiation yields. This groundbreaking experiment combining both the superposition of the forward and backward spin currents and the Fabry-Perot effect makes W/CoFeB/Pt trilayer heterostructures turn out to have even higher efficiency than those from commercial nonlinear crystals and photoconductive antennas when driven by a 10 fs Ti:sapphire laser oscillator [2]. In this case, the ultrashort pumping pulse constrains the efficiency from nonlinear crystals because optical rectification radiation mechanism requires appropriate effective interaction length which is strongly correlated to the pumping pulse duration [19]. For photoconductive antennas, the ultrahigh peak intensity due to ultrashort pulses suppresses the efficiency because the tolerance of pumping power limits the efficiency caused by the antenna breakdown [19]. However, for conventional ultrafast laser systems widely used in most laboratories, the pulse durations are always larger than 10 fs. Hence it is still very challenging to obtain higher or even comparable efficiencies in the spintronic emitters than those from nonlinear crystals or commercial photoconductive antennas. There is still plenty room for improvement for longer pulse pumping.

In this work, we directly choose the well demonstrated W/CoFeB/Pt trilayer heterostructures with high optical-to-terahertz energy conversion efficiency [2], and systematically study and verify the possibility of using remnant magnetization for stable terahertz emission excited by a Ti:sapphire laser oscillator with 100 fs pulse duration. Through deeply investigating the impact of pumping laser wavelength on the radiated terahertz strength and properties, we find that variation of pumping wavelength has strong correlations on the terahertz emission performance, which is not consistent as that has been reported in [21]. Besides, the incidence angle-resolved terahertz emission, time-domain reflection and transmission spectroscopy, and infrared absorption are systematically measured. Experimental and qualitatively theoretical results illustrate that the terahertz emission can be scaled up with oblique incidence. Our investigations not only help further deeply understanding the femtosecond spin dynamics, but also benefit for building high performance spintronic terahertz sources and terahertz opto-spintronic devices.

## II. Experimental Setup and Sample Preparation

The experimental setup is schematically illustrated in Figure 1a. We employ a home-made incidence angle-resolved terahertz time-domain spectrometer to measure the terahertz emission and time-domain spectroscopy with transmission and reflection modes. The system is driven by a commercial Ti:sapphire laser oscillator (MaiTai, Spectra-Physics) with variable central wavelength from 710 nm to 860 nm in horizontal polarization. The pulse duration is 100 fs and the repetition rate is 80 MHz. 90% of the pumping power is collimated and focused with two parabolic mirrors (P1 and P2 in Figure 1a) onto the magnetic nanofilm which is mounted on a remotely controlled sample holder whose rotation angles can be varied from 0° to 70° for the transmission mode while 10°-70° for the reflection mode. The transmitted terahertz signal is collimated and then focused onto 1 mm thick ZnTe detector together with the probing infrared beam for electro-optic sampling which is optimized for terahertz waves with horizontal polarization. The reflected terahertz signal is first collected by a parabolic mirror (P5), and the collimated reflection signals are guided through five aluminum mirrors (M1-M5) to the transmission optical path. P5 and M1 are synchronized together with the rotational sample holder. When M5 is flipped up, it is for the transmission measurement, while down for the reflection mode. The systems can be easily changed for terahertz time-domain spectroscopy measurement when we insert a commercial photoconductive antenna (Zomega terahertz) at the focus of the pumping beam before P1. The system is purged by dry nitrogen to rule out the influence from water vapor, and all the measurements are carried out at room temperature. The spintronic terahertz emission examined in our experiments is from a W/CoFeB/Pt trilayer heterostructure fabricated on 1 mm thick glass substrate using magnetron sputtering method [20]. Each layer has the same thickness of 1.8 nm. The sample was first externally magnetized with 50 mT magnetic field, and then for the terahertz emission and time-domain spectroscopy, there is no more magnets around it.

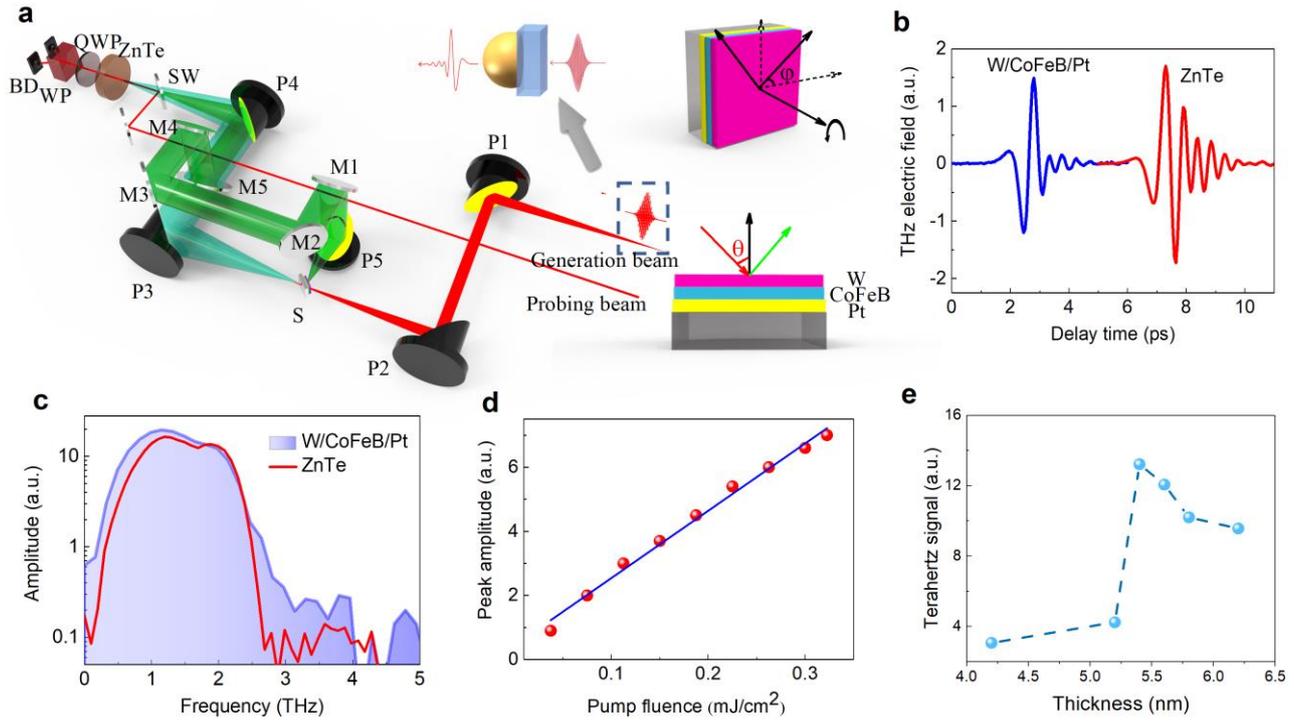

**Fig. 1 | Comparable achievement obtained from W/CoFeB/Pt through angle-resolved transmission and reflection terahertz emission spectroscopy. a,** Schematic diagram of the experimental setup. P1-4: 90° off-axis parabolic mirrors; M1-5: aluminum reflection mirrors; SW: silicon wafer for combing the probing beam together with terahertz waves; S: sample of W/CoFeB/Pt with 1.8 nm thickness for each layer; QWP: quarter wave plate; WP: wollaston prism; BD: balanced detector. $\varphi$ is the azimuthal angle of the sample, while $\theta$ is the incidence angle of the pumping beam. **b, and c,** Terahertz temporal waveforms from W/CoFeB/Pt and ZnTe, and their corresponding Fourier transformed spectra with $\varphi=\theta=0^\circ$. **d,** Terahertz emission amplitude for the transmission measurement mode as a function of the pump fluence in W/CoFeB/Pt trilayers. **e,** Dependence of the sample thickness on the radiated terahertz peak-to-peak signal.

We first show the best results detected in the W/CoFeB/Pt sample with 5.4 nm total thickness under the maximum pump fluence of ~0.3 μJ/cm$^2$ with oblique incidence of 45° at optimal central wavelength pumping of 790 nm. We obtain a magnitude in the sample comparable with that from a 1 mm thick ZnTe emitter, as shown in Figure 1b. The corresponding Fourier transformed spectra are plotted in Figure 1c, in which we achieve a slightly broader spectrum and a little higher efficiency in lower frequency region <2 THz. The terahertz yields could be further scaled up, as shown in Figure 1d, since the peak amplitude obeys a linear increase behavior as a function of the pump fluence. All the features demonstrate the feasibility of making spintronic terahertz sources very promising.

We also examined five trilayer samples with total thickness of 4.2, 5.2, 5.6, 5.8 and 6.2 nm, respectively. No matter the total thickness is smaller or larger than 5.4 nm, the detected terahertz signals are reduced, which agrees very well with the reported experimental results [2]. The mechanism here is due to the competition behavior between the reflective losses at the cavity faces and the transmission losses in the bulk metals. When the total thickness of the sample is less than the optimal thickness (in our case, 5.4 nm), the number of reflections and the corresponding losses at the FM/HM interfaces decrease with the increase of the total thickness of the sample, leading to the enhanced terahertz emission. When the sample thickness exceeds the optimal thickness, the transmission losses in the metals increase, resulting in the reduction of the terahertz radiation. Following parts will be focused on how we achieve this result and manifest the optimization procedures and their physical mechanism.

## III. Results and Discussions
## A. Verification of the external magnetization proposal

Our first demonstration is to corroborate the possibility of using remnant magnetization for our spintronic emitters. In this case, the pumping fluence is fixed at 0.3 μJ/cm$^2$ to avoid saturation behavior probably due to laser heating-induced conductivity variation or spin accumulation effect [4]. Because the electro-optic sampling in ZnTe crystal is optimized for detecting horizontally polarized terahertz pulses, when we rotate the magnetized sample from 0° to 180°, as exhibited in Figure 2a, the emitted terahertz polarity is reversed. However, when further increasing the azimuthal angle to 360°, the terahertz phase returns to its pristine state (see Figure 2a and b). This confirmatory experiment manifests the inverse spin Hall effect as the radiation mechanism, with the equation of

$$\vec{J}_c = \gamma \vec{J}_s \times \left[ \vec{M} / \left| \vec{M} \right| \right] \tag{1}$$

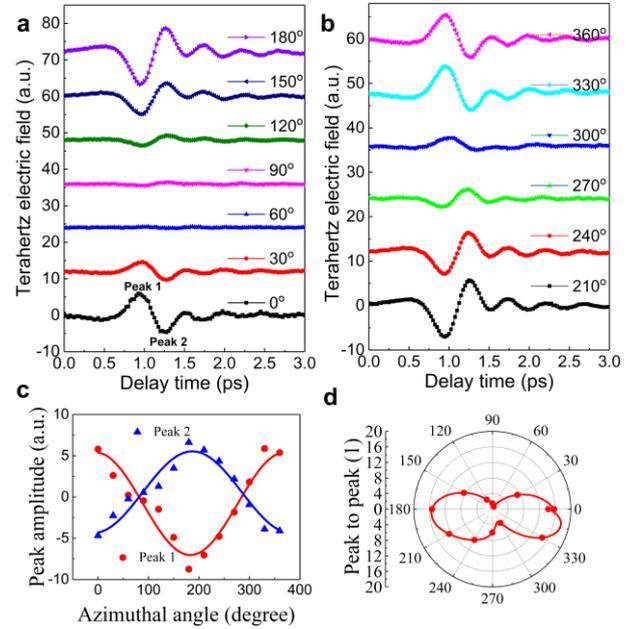

**Fig. 2 | Verification of the feasibility of external magnetization scheme. a and b,** The measured terahertz transmitted emission for arbitrary azimuthal angles. **c,** The two peak values of the measured terahertz signals. **d,** Dependence of the peak to peak value of terahertz signals on the azimuthal angle.

where $\vec{J}_c$ is the spin converted transverse charge current, $\gamma$ is the spin Hall angle for W and Pt, $\vec{J}_s$ is the light induced spin current, and $\vec{M}$ is the magnetization. The emission amplitude pattern as a function of the azimuthal angle is summarized in Figure 2c and d. Compared with the signals from the azimuthal angles of 0° and 360°, the repeatability of the emission strength and polarity demonstrates the proposed external magnetization method is feasible [1].

### B. Optimization of pumping central wavelength

To gain optimal pumping central wavelength and obtain a deep insight into the optimization mechanism, we fix the sample azimuthal angle at 0°. Firstly, we examine the input pumping spectra as well as its corresponding output spectra when varying the central wavelength from the laser oscillator, see typical spectra in Figure 3a. There is no remarkably difference between the input and output spectra, except the spectrum amplitude decreases by ~62%. This proves that the material has no distinct absorption signatures in this infrared frequency region, which implies that the laser absorption induced spin current generation for all the infrared wavelengths is almost the same. Figure 3b gives the inverse Fourier transformed pulse widths corresponding to the input and output spectra for different central wavelengths. They preserve the same profiles which implies that the generation of terahertz waves has no significant modulation on the pumping pulse duration. However, for different central wavelengths, the useful average powers for terahertz generation and detection are different (see Figure 3d), which is limited by the pumping laser oscillator. Figure 3c depicts the emitted terahertz transients from which we find that the terahertz waveform obtained at 790 nm appears earlier than those for other central wavelengths. The possible mechanism may be attributed to the dispersion of ZnTe detection crystal, but it is probably not the predominate effect because of the observed time-domain symmetrical distribution behavior from 750 to 815 nm. Therefore, the other possible reason may be entangled with the variation of the permittivity due to change of pump fluence in the heterostructure emitter, which is ill-defined and needs further deep investigations.

Figure 3h depicts the terahertz peak-to-peak signals normalized by both of the pump and probe powers. Surprisingly, it is obvious that the optimal central wavelength is around 790 nm. The terahertz signal is three times higher than that measured at 755 nm, which is different from the experimental phenomena observed in reference [21]. The possible mechanism as proposed in reference [2] was due to the Fabry-Perot effect similar to the observed variations of the sample thickness-dependent terahertz yields in CoFeB/Pt bilayers. Firstly, for the pumping laser frequency range, the refractive index for W and Pt is ~2-4, while the exactly value for CoFeB is not known. When the total thickness is 5.4 nm in our case, the first round-trip of 10.8 nm between W-glass interface and Pt-air interface enables the phase shift <20°. Likewise, considering this effect at 1 THz, the optical path of terahertz waves is 1-2 μm resulting in less than 2° phase shift. Namely it is noticeable not due to the Fabry-Perot effect of both the pumping laser and the generated terahertz waves. As we have explained in the thickness dependent terahertz emission in Figure 1e, the central wavelength dependent terahertz optimization is also due to the competition between transmission and reflection losses in the samples [22]. Our sample thickness (5.4 nm) is optimized for the central wavelength at 790 nm, hence the effect of changing the pumping central wavelength is equivalent to varying sample thickness. Likewise, the essentially flat dependence of the excitation central wavelength on the terahertz signals observed by Herapath et al. [21], can also be well explained by the aforementioned effect. The sample thickness used in their work was 5.8 nm which was optimized for the central wavelength at ~800 nm [2]. According to our observed almost symmetrical distribution of the terahertz signals as a function of the excitation central wavelength, it is very possible to be expected that the pumping wavelength dependence would become essentially flat when the central wavelength is spanning from 900 nm to 1500 nm, which does not include the optimal wavelength of 800 nm.

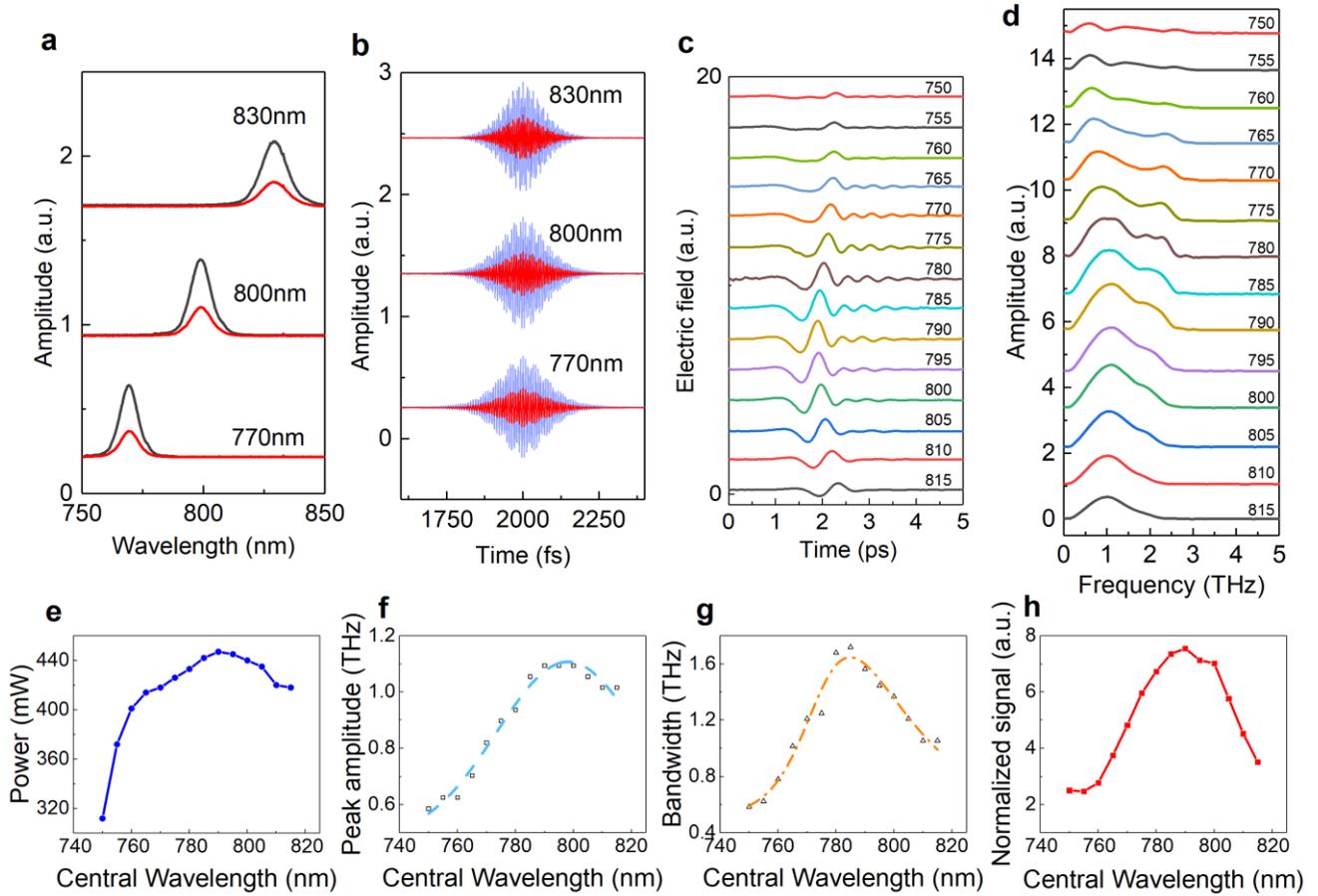

**Fig. 3 | Pumping central wavelength-dependent terahertz emission. a,** The input spectra for the central wavelengths at 770, 800, and 830 nm (black solid lines), and their corresponding output spectra after terahertz generation (red solid lines). **b,** The inverse Fourier transformed pulse durations for input (blue) and output (red) spectra. **c and d,** Measured terahertz temporal waveforms and their corresponding spectra as a function of the pumping central wavelengths. **e,** Pumping average power from the laser oscillator for different central wavelengths. **f and g,** The radiated terahertz peak frequency and the spectrum range as a function of the central wavelength. **h,** Terahertz signals normalized to the excitation and detection average power as a function of the central wavelengths.

As illustrated in Figure 3d, we can tune the emitted terahertz peak frequency as well as the frequency range by varying the pumping central wavelength. As summarized in Figure 3f, when increasing the central wavelengths from 750 nm to 815 nm, the radiated terahertz peak frequency first increases from 0.6 THz to 1.1 THz with the highest frequency appearing at 790 nm, and then decreases back to 1.0 THz. Following almost the same tendency, the frequency range (defined as the full width at half maximum of the terahertz spectra) also broadens itself from 0.6 THz to 1.7 THz (785 nm) and then reduces back to 1.1 THz (see Figure 3g). The tunability mechanism may be due to the variable permittivity for different terahertz wavelengths. For the reflection losses at the heterostructure interfaces and the transmission losses in the metal bulks, different terahertz wavelengths have different refractive indices and corresponding absorption coefficients. However, for this kind of thin films with only several nanometer thicknesses, it is very challenging to genuinely obtain the complex permittivity and consequently the complex refractive index. Therefore, there exist optimal sample thicknesses (equal to variation of the excitation central wavelengths) for particular terahertz frequencies. A straightforward summary can be formed that different pumping central wavelengths should correspond to various optimal thickness and it is very important to achieve highly efficient terahertz emission in FM/NM heterostructures excited by femtosecond laser pulses.

### C. Dependence on the oblique incidence

To further explore the scaling up possibility of the terahertz generation efficiency, we vary the incidence angle of the pumping laser with respect to the sample surface. Figure 4 summarizes the experimental results of the measured maximum transmitted terahertz yield which appears at ~45°, and it starts to reduce afterwards due to the sample holder blocking. For the optimal incidence angle, the peak-to-peak terahertz fields have been enhanced by 100%. Similar behavior is also observed for the reflection mode measurements, and the detected terahertz signals increase roughly monotonously (Figure 4b). From 15° to 60°, the terahertz fields are scaled up by ~200%. The observed behavior may be attributed to the following mechanisms of incidence angle-dependent infrared light absorption enhancement, or the increase of the terahertz out-coupling due to Brewster angle effect for the emitting sample.

To elucidate the mechanism, Figure 5a exhibits the W/CoFeB/Pt absorptance as a function of the incidence angle under illumination of 790 nm central wavelength.

The absorptance increases till 20°, and then decreases monotonously till 50° (beam blocking by sample holder). This observation can be used to rule out the first possible mechanism. To unveil the reduced Fresnel reflection of W/CoFeB/Pt due to Brewster angle effect, we measured the incidence angle-dependent transmitted terahertz signals as well as their peak-to-peak amplitudes which are summarized in Figure 5c and b, respectively. The incidence angle dependent reflected terahertz signal decreases while the roughly linear increase behavior for the transmission signals proves phenomenologically the high reasonability of the second speculate. We also measured the reflected terahertz signals dependent on the incidence angle (see Figure 5b). We use the transmitted terahertz signal at normal incidence to extract the effective terahertz permittivity of W/CoFeB/Pt on glass substrate. Figure 6 shows the calculated effective refractive index, which is almost the same as that from bare glass substrate. The total thickness of W/CoFeB/Pt trilayer is 5.4 nm in our case, which seems to have no big influence on the terahertz wave interaction. The absorption coefficient given in Figure 6b demonstrates that the glass substrate has heavy absorption for terahertz waves which can also be further improved in the optimization of the substrates. Figure 6d is derived from Figure 6c, and the enhancement of the transmittance together with the experimentally observed (see Figure 5b) and the calculated decreasing of the reflected signals (see Figure 6c) demonstrates the reasonable correspondence between the scaling up behavior of the radiated terahertz waves and the incidence angle. The enhanced reflection emission when increasing the incidence angle is still not clear and needs further deeply investigations.

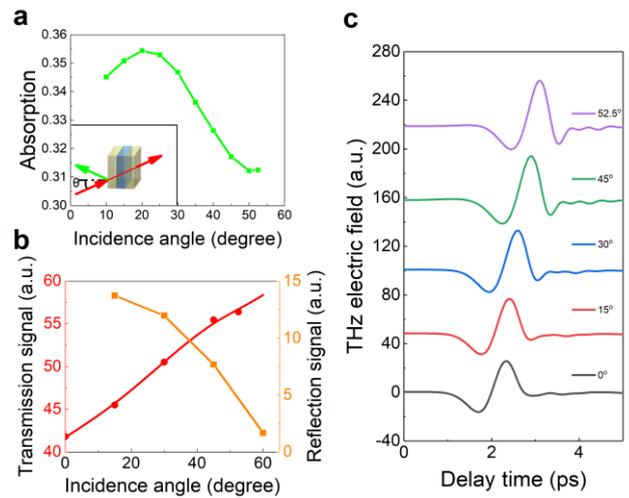

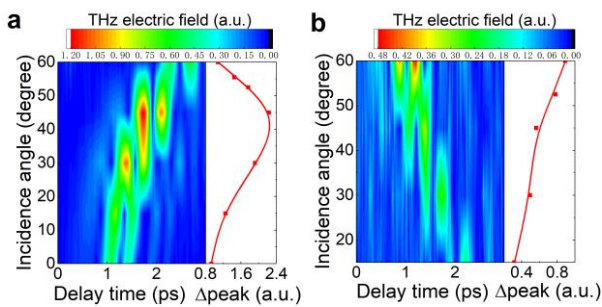

**Fig. 4 | Incidence angle optimization for the terahertz emission. a, and b,** The measured terahertz electric field amplitudes in the W/CoFeB/Pt trilayer measured in transmission and reflection modes, respectively.

**Fig. 5 | Verification of the possible terahertz enhancement mechanism for oblique incidence. a,** Incidence angle dependence of absorption for the pumping light. The inset shows this experimental setup. **b,** Dependence of peak to peak transmitted and reflected terahertz electric fields on the incidence angle. **c,** The measured terahertz transmitted signals for the incidence angles of 0°, 15°, 30°, 45°, and 52.5°.

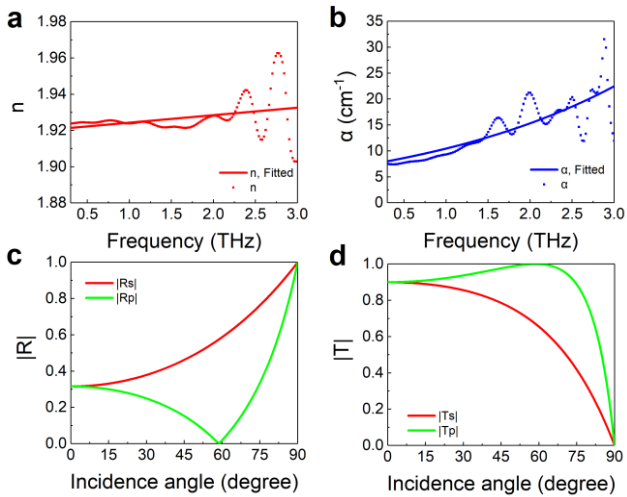

**Fig. 6 | Calculation of the terahertz refractive index and absorption coefficient of W/CoFeB/Pt/Glass trilayer. a, and b,** The effective refractive index and absorption coefficient of the W/CoFeB/Pt/Glass, respectively. The dot lines are experimental results while solid lines for fitting. **c,** Dependence of the reflectivity of s and p polarized terahertz waves on the incidence angle, respectively. **d,** Incidence angle-dependent terahertz transmittance of s and p polarization.

### IV. Conclusion

In summary, we demonstrate the feasibility of external magnetization scheme can be used to generate stable terahertz radiation in W/CoFeB/Pt heterostructures pumped by nJ Ti:sapphire laser oscillator. We find that there exists an optimal sample thickness for a particular excitation central wavelength, and an enhancement behavior for both transmitted and reflected terahertz radiation under oblique incidence. Through strenuously optimizing the sample thickness, pumping laser central wavelength and the incidence angle, we finally obtain a highly efficient terahertz source comparable with commercial nonlinear crystals under 100 fs pulse duration excitation, in which optical rectification induced terahertz efficiency has been optimized. In the future, we envision to further develop cost-effective, highly efficient, ultrabroadband, and tunable intense spintronic terahertz sources based on the currently verified principles. We believe our investigations will have considerable interest for the perspective of terahertz spintronic devices and platforms.

### Acknowledgements

This work is supported by the National Nature Science Foundation of China (Grants No. 11827807, 11520101003, 11861121001, 61831001 and 61775233), the Strategic Priority Research Program of the Chinese Academy of Sciences (Grant No. XDB16010200 and XDB07030300), the National Basic Research Program of China (Grant No. 2014CB339800), the International Collaboration Project (Grant No. B16001), and the National Key Technology Program of China (Grant No. 2017ZX01032101). Dr. Xiaojun Wu thanks the "Zhuoyue" Program and "Qingba" Program of Beihang University (Grant No. ZG216S1807, ZG226S1832, and KG12052501). Dr. Tianxiao Nie thanks the support from the 1000-Young talent program of China.

### Conflict of Interest

The authors declare no competing financial interest.